\documentstyle[sprocl]{article}

\def\al{\alpha}
\def\be{\beta}

\def\et{\eta}

\def\ka{\kappa}

\def\cL{{\mathcal L}}

\def\fr#1#2{{{#1} \over {#2}}}
\def\half{{\textstyle{1\over 2}}}

\def\frac#1#2{{\textstyle{{#1}\over {#2}}}}

\def\vev#1{\langle {#1}\rangle}

\def\lsim{\mathrel{\rlap{\lower4pt\hbox{\hskip1pt$\sim$}}
    \raise1pt\hbox{$<$}}}
\def\gsim{\mathrel{\rlap{\lower4pt\hbox{\hskip1pt$\sim$}}
    \raise1pt\hbox{$>$}}}
\def\sqr#1#2{{\vcenter{\vbox{\hrule height.#2pt
         \hbox{\vrule width.#2pt height#1pt \kern#1pt
         \vrule width.#2pt}
         \hrule height.#2pt}}}}

\def\etal{{\it et al.}}

\def\pt#1{\phantom{#1}}

\def\vb#1#2{e_{#1}^{{\pt{#1}}#2}}

\newcommand{\beq}{\begin{equation}}
\newcommand{\eeq}{\end{equation}}
\newcommand{\bea}{\begin{eqnarray}}
\newcommand{\eea}{\end{eqnarray}}
\newcommand{\bit}{\begin{itemize}}
\newcommand{\eit}{\end{itemize}}

\begin{document}

\title{SPONTANEOUS LORENTZ VIOLATION, NAMBU-GOLDSTONE
MODES, AND MASSIVE MODES}

\author{ROBERT BLUHM}

\address{Department of Physics and Astronomy, Colby College\\
Waterville, ME 04901, USA}

\bigskip

\maketitle

\abstracts{
In any theory with spontaneous symmetry breaking,
it is important to account for the massless Nambu-Goldstone
and massive Higgs modes.
In this short review,
the fate of these modes is examined for the case
of a bumblebee model,
in which Lorentz symmetry is spontaneously broken.
}

\bigskip

\section{Spontaneous Lorentz violation}   

Spontaneous symmetry breaking
has three well-known consequences.
The first is Goldstone's theorem,
which states that massless Nambu-Goldstone (NG)
modes should appear when a continuous symmetry
is spontaneously broken.
The second is that massive Higgs modes can appear,
consisting of excitations out of the degenerate minimum.
The third is the Higgs mechanism,
which can occur when the broken symmetry is local.

In conventional particle physics,
these processes involve
a scalar field with a nonzero
vacuum expectation value
that induces spontaneous breaking of
gauge symmetry.
However, in this paper,
these processes are examined for the case where it
is a vector field that has a nonzero vacuum value
and where the symmetry that is spontaneously
broken is Lorentz symmetry.

The idea of spontaneous Lorentz violation is important in
quantum-gravity theory.
For example, in string field theory,
mechanisms have been found that can lead to
spontaneous Lorentz violation.\cite{ks}
It has also been shown that
spontaneous Lorentz breaking is
compatible with geometrical consistency conditions
in gravity,
while explicit Lorentz breaking is not.\cite{akgrav}

The Standard-Model Extension (SME)
describes Lorentz-violating interactions at the level of effective
field theory.
\cite{sme,rbsme}
In recent years,
numerous tests of Lorentz symmetry have been performed,
and sensitivities to extremely small effects has been attained.\cite{aknr}.

\vfill\eject

\section{Bumblebee models}

The simplest example of a theory with spontaneous
Lorentz violation is a bumblebee
model.\cite{ks,akgrav,bj,yn,kl01,baak05,rbak,rbffak,bgpv,cdgt,ms,cfjn,ap}
It is defined as an effective field theory with a
vector field $B_\mu$ that has
a nonzero vacuum value, $\vev{B_\mu} = b_\mu$.
In general, the theory includes gravity with a metric field $g_{\mu\nu}$.
The interactions in the theory can be written in terms of a potential $V$,
which has a minimum when $B_\mu$ and $g_{\mu\nu}$
equal their vacuum values.

Bumblebee models are invariant under both diffeomorphisms
and local Lorentz transformations.
To reveal the local Lorentz symmetry,
a vierbein formalism is used.
The vierbein $\vb \mu a$ relates tensor fields
defined on the spacetime manifold,
e.g., $g_{\mu\nu}$ and $B_\mu$,
to the corresponding fields defined in
a local Lorentz frame,
e.g., $\et_{ab}$ and $B_a$.

Spontaneous Lorentz violation occurs when the vector
field has a nonzero vacuum solution,
$b_a$,
in a local Lorentz frame.
Since the metric and vierbein have background
values in the spacetime frame,
the bumblebee vector in the spacetime frame
also has a nonzero vacuum value,
$b_\mu = \vev{\vb \mu a}  \, b_{a} $,
which therefore spontaneously breaks diffeomorphism symmetry.
This illustrates a general result
that whenever local Lorentz symmetry
is spontaneously broken by a local background tensor,
then diffeomorphism symmetry is spontaneously broken as well
\cite{rbak}.

A variety of bumblebee models can be defined
depending on the definitions of the potential $V$
and kinetic terms for $g_{\mu\nu}$ and $B_\mu$.
If an Einstein-Hilbert action is chosen for the
gravitational sector,
then the excitations of the metric
include the usual graviton modes as
in general relativity.
The NG modes
can be identified as virtual symmetry transformations
about the vacuum solution
that stay in the
minimum of the potential $V$,
while the massive modes are excitations that move
out of the minimum.

The interpretation of the NG modes depends on the
choice of kinetic terms for the bumblebee field.
In one class of models,
the bumblebee field is treated as a vector in
a vector-tensor theory of gravity,
and a Will-Nordtvedt form of kinetic term is used.\cite{cw}
In this scenario,
the NG modes are considered additional gravitational excitations.
An alternative interpretation is to treat the field $B_\mu$
as a generalized vector potential.
In this case,
a Maxwell kinetic term is used,
and the NG modes are interpreted as photons.\cite{bj,yn,rbak,rbffak}

The original model of Kosteleck\'y and Samuel (KS)
uses a Maxwell kinetic term for $B_\mu$ and a
potential $V$ that has a minimum when the
condition $B_\mu B^\mu = \pm b^2$ holds.\cite{ks}
The KS bumblebee model in a Minkowski background
has Lagrangian
\beq
\cL_{\rm KS} = \fr 1 {16 \pi G} R
- \fr 14  B^{\mu\nu} B_{\mu\nu}
- V + B_\mu J^\mu + \cL_M .
\label{KSBB}
\eeq
Here, $B_{\mu\nu} = D_\mu B_\nu - D_\nu B_\mu$,
is the generalized field strength
defined using gravitational covariant derivatives,
$J^\mu$ is a matter current,
and $\cL_M$ is the Lagrangian for the matter fields.
A generic form for the potential $V$
is a power-series expansion
in $B_\mu B^\mu$,
which when truncated
to second order is
\beq
V = \half \ka (B_\mu B^\mu \pm b^2)^2 ,
\label{Vkappa}
\eeq
where $\ka$ is a constant.

A important feature of all bumblebee models,
including the KS model,
is that they do not have local $U(1)$ gauge symmetry,
which is broken explicitly by the potential $V$.
However,
with conventional matter couplings a global U(1)
symmetry can exist.
This results in conservation of charge in the matter sector,
as given by the equation, $D_\mu J^\mu = 0$.
This conservation law also ensures that the bumblebee model is
perturbatively stable about the minimum solution.\cite{bgpv}

\section{Nambu-Goldstone and massive modes}

The field equations are obtained by varying
$\cL_{\rm KS}$ with respect to $g_{\mu\nu}$ and $B_\mu$.
These can be used to determine the propagating
gravitational, NG, and massive modes.\cite{rbak,rbffak}
The excitations of the metric about the vacuum
can be written as
$h_{\mu\nu} = g_{\mu\nu} - \et_{\mu\nu}$,
which includes the usual graviton modes.
A convenient gauge choice for decoupling
the NG modes is the condition
$b^\mu h_{\mu\nu} = 0$.
Imposing it allows the bumblebee field to be written as
\beq
B_\mu = b_\mu + A_\mu + \hat b_\mu \be ,
\label{expandB}
\eeq
where $\hat b$ is a unit vector along the direction of $b_\mu$.
In this expansion,
$A_\mu$ is transverse to $b_\mu$,
obeying the
condition $b^\mu A_\nu = 0$,
while $\be$ is along the direction of $b_\mu$.

The excitations $A_\mu$ are the NG modes
generated by the broken Lorentz transformations.
For these excitations, $V$ remains in the minimum
of the potential.
The condition $b^\mu A_\nu = 0$ has the form of
an axial gauge-fixing constraint.
Of the three remaining components in $A_\mu$,
two propagate like photons in a fixed axial gauge,
while the third is an auxiliary field due to a constraint
in the form of a modified version of Gauss' law.

The excitation $\be$ along the direction of $\hat b_\mu$
is a massive-mode (or Higgs excitation).
It can be written as
\beq
\be = \fr {\hat b^\mu  (B_\mu - b_\mu)} {\hat b^\al  \hat b_\al} .
\label{beta}
\eeq
These excitations do not stay in the potential minimum.
At leading order,
the massive mode $\be$ does not propagate as a free field.
Instead,
it acts as a background source of both charge and energy density.
\cite{rbak,rbffak}
The massive mode in the KS bumblebee model
is an extra degree of freedom compared to Einstein-Maxwell theory.
\cite{bgpv}
Its behavior depends on the initial conditions that are specified for it.
In particular,
it has been shown that
if the massive mode vanishes initially, it vanishes for all time,
and the KS bumblebee model reduces to Einstein-Maxwell theory
with photons as NG modes.
If instead, excitations of the massive mode $\be$ are allowed,
then an alternative theory to Einstein-Maxwell theory results,
which could be of interest in attempts to find explanations of
dark matter or dark energy.

For example,
it is found that both the electromagnetic and gravitational
potentials for a point particle with mass $m$ and charge $q$
are modified by the presence of the massive mode.
The specific forms of the modified potentials depend on the
initial value of the massive mode,
and there are therefore numerous cases that could be explored.
However,
in the large-mass limit
(e.g., approaching the Planck scale),
excitation of the massive mode is highly suppressed,
and the static potentials approach the conventional Coulomb
and Newtonian forms.
As a result,
it is found that the usual Einstein-Maxwell solutions
(describing both propagating photons and the usual static potentials)
can emerge from the KS bumblebee model,
despite the absence of local U(1) gauge invariance.

\section{Gravitational Higgs mechanisms}

Since in the context of gravity,
Lorentz symmetry is a local symmetry,
the possibility of a Higgs mechanism occurs as well.
However, with two sets of broken symmetries,
local Lorentz transformations and diffeomorphisms,
there are actually two types of  Higgs mechanisms
to consider.

In the original paper by Kosteleck\'y and Samuel,
it was shown that a conventional Higgs mechanism
involving the metric does not occur.\cite{ks}
This is because the quadratic term that is generated by covariant derivatives
involves the connection,  which consists of derivatives of the metric
and not the metric itself.
As a result,
no mass term for the metric is generated
according to the usual Higgs mechanism.

However, more recently, it has been shown that a
Higgs mechanism can also
occur that involves the spin connection, \cite{rbak,rbffak}
which appears in covariant derivatives acting on local tensor components.
When a local tensor has a vacuum value,
quadratic mass terms for the spin connection can be generated.
However, a viable Higgs mechanism can
occur only if the spin connection is a dynamical field.
This requires that there is torsion in the theory
and that the geometry is Riemann-Cartan.
As a result,
a conventional Higgs mechanism for the spin connection is possible,
but only in a Riemann-Cartan geometry.
If torsion is permitted,
a number of new types of models with Lorentz breaking can be explored.
However, finding a viable model free of ghosts
remains a challenging and open issue.

\section*{Acknowledgments}

This work was supported
by NSF grant PHY-0854712.

\end{document}